\documentclass[12pt]{article}
\usepackage{amsfonts}

\usepackage{graphicx}
\usepackage{amsmath}
\usepackage{float}

\textwidth=6.5in \hoffset=-.75in \voffset=-1in \numberwithin{equation}{section}
\numberwithin{figure}{section}

\begin{document}

\begin{titlepage}
\vspace{1cm}
\begin{center}
{\Large \bf {Atiyah-Hitchin in Five Dimensional Einstein-Maxwell Theory}}\\
\end{center}
\vspace{2cm}
\begin{center}
A. M. Ghezelbash{ \footnote{ E-Mail: masoud.ghezelbash@usask.ca}}
\\
Department of Physics and Engineering Physics, \\ University of Saskatchewan, \\
Saskatoon, Saskatchewan S7N 5E2, Canada\\
\vspace{1cm}
\vspace{2cm}
\end{center}

\begin{abstract}
We construct exact solutions to five-dimensional Einstein-Maxwell theory based on Atiyah-Hitchin space. The solutions cannot be written explicitly in a closed form, so their properties are investigated numerically. The five-dimensional metric is regular everywhere except on the location of original bolt in four-dimensional Atiyah-Hitchin base space. On each time-fixed slices, the metric, asymptotically approaches an Euclidean Taub-NUT space.

\end{abstract}
\end{titlepage}\onecolumn 
\bigskip 

\section{Introduction}

The Atiyah-Hitchin space is a part of the set of two monopole solutions of
Bogomol'nyi equation. The moduli space of solutions is of the form%
\begin{equation}
\mathbb{R}^{3}\otimes \frac{S^{1}\otimes \mathcal{M}}{\mathbb{Z}_{2}},
\end{equation}%
where the factor $\mathbb{R}^{3}\otimes S^{1}$ describes the center of mass
of two monopoles and a phase factor that is related to the total electric
charge of the system. The interesting part of the moduli space is the 
four-dimensional manifold $\mathcal{M}$, which has self-dual curvature. The
self-duality comes from the hyper-K\"{a}hler property of the moduli space.
Since $\mathbb{R}^{3}\otimes S^{1}$ is flat and decouples from $\mathcal{M}$%
, the four-dimensional manifold $\mathcal{M}$ should be hyper-K\"{a}hler,
which is equivalent to a metric with self-dual curvature in four dimensions.
The manifold $\mathcal{M}$ describes the separation of the two monopoles and
their relative phase angle (or electric charges). A further aspect
concerning $\mathcal{M}$ is that it should be $SO(3)$\ invariant, since two
monopoles do exist in ordinary flat space; hence the metric on $\mathcal{M}$
can be expressed in terms of three functions of the monopole separation.
Self-duality implies that these three functions obey a set of first-order
ordinary differential equations.
This space has been used recently for construction of five-dimensional three-charge supergravity solutions that only have a rotational $U(1)$
isometry \cite{B4} as well as construction of M-brane solutions \cite{GH}. 
Moreover, Atiyah-Hitchin space and its various generalizations were identified
with the full quantum moduli space of $\mathcal{N}=4$ supersymmetric gauge
theories in three dimensions \cite{seib}.

Moreover, in the context of string theory and brane world, investigations on 
black hole (ring) solutions in higher dimensions have attracted a lot of attention.
It is believed that in the strong coupling limit, many horizonless
three-charge brane configurations undergo a geometric transition and become
smooth horizonless geometries with black hole or black ring charges \cite{B1}. These charges come completely from fluxes wrapping on non-trivial cycles.
The three-charge black hole (ring) systems are dual to the states of
corresponding conformal field theories: in favor of the idea that non-fundamental-black hole
(ring) systems effectively arise as a result of many horizonless
configurations \cite{B2,Ma1}. 
In eleven-dimensional supergravity, there are solutions based on
transverse four-dimensional hyper-K\"{a}hler metrics (which are equivalent to metrics
with self-dual curvatures). The hyper-K\"{a}hlericity of transverse metric guarantees 
(at least partially) to have supersymmetry \cite{G1}. There are also
many solutions to five-dimensional minimal supergravity. In
five-dimensions, unlike the four dimensions that the only horizon topology
is 2-sphere, we can have different more interesting horizon topologies such
as black holes with horizon topology of 3-sphere \cite{Myers}, black rings
with horizon topology of 2-sphere $\times $ circle \cite{Em1,Em2}, black
saturn: a spherical black hole\ surrounded by a black ring \cite{El1}, black
lens which the horizon geometry is a Lens space $L(p,q)$ \cite{Ch1}. All
allowed horizon topologies have been classified in \cite{Ca1,He1,Ga1}.
Recently, it was shown how a uniqueness theorem might be proved for black
holes in five dimensions \cite{Ho1,Ho2}. It was shown stationary,
asymptotically flat vacuum black holes with two commuting axial symmetries
are uniquely determined by their mass, angular momentum and rod structure.
Specifically, the rod structure \cite{Ha1} determines the topology of
horizon in five dimensions.
In the references \cite{Ishi1, Ishi2, Ishi3, Ishi4}, the authors constructed (multi) black hole solutions in the five-dimensional Einstein-Maxwell theory (with and without cosmological constant) based on four-dimensional Taub-NUT and Eguchi-Hanson spaces. Both spaces have self-dual curvatures and can be put into a Gibbons-Hawking form. Although hyper-K\"{a}hler Atiyah-Hitchin space also has self-dual curvature, it cannot be put into a Gibbons-Hawking form.

Motivated by these facts, in this article we try to construct solutions to five-dimensional Einstein-Maxwell theory based on Atiyah-Hitchin space. We note that
hyper-K\"{a}hler Atiyah-Hitchin geometries (unlike Gibbons-Hawking geometries) do not have any tri-holomorphic $U(1)$ isometry, hence our solutions could be used to study the 
physical processes that do not respect any tri-holomorphic $U(1)$ symmetry.
We consider the Atiyah-Hitchin space as a base space and by performing a coordinate transformation, we are able to considerably simplify the structure of the five-dimensional metric. This is the first step toward construction of more sophisticated solutions (such as black holes or rings) in five-dimensional Einstein-Maxwell theory based on Atiyah-Hitchin space.

The outline of this paper is as follows. In section \ref{sec:5Dreview},
we review briefly the Einstein-Maxwell theory, the Atiyah-Hitchin space and its features. In section \ref{sec:sol},
we present our solutions based on two forms for the Atiyah-Hitchin space and discuss the 
asymptotics  of solutions. We conclude in section \ref{sec:con} with a summary of our solutions and possible future research directions.

\section{Five-dimensional Einstein-Maxwell Theory and Atiyah-Hitchin Space}

\label{sec:5Dreview}

The five-dimensional Einstein-Maxwell theory is described by the action
\begin{equation}
S=\frac{1}{16\pi}\int d^5x\, \sqrt{-g}(R-F_{\mu\nu}F^{\mu\nu}),
\end{equation}
where $R$ and  $F_{\mu\nu}$ are five-dimensional Ricci scalar and Maxwell field.
The Einstein and Maxwell equations are
\begin{eqnarray}
R_{\mu\nu}-\frac{1}{2}Rg_{\mu\nu}&=&2F_{\mu\lambda}F^{\nu\lambda}-\frac{1}{2}g_{\mu\nu}F^2,\label{EEQ} \\
F^{\mu\nu}_{;\nu}&=&0,\label{GEQ}
\end{eqnarray}
respectively. We take the following form for the five-dimensional metric
\begin{equation}
ds_5^{2}=-H(r)^{-2}dt^{2}+H(r)ds_{AH}^2,
\label{ds5}
\end{equation}%
and the only non-vanishing component of gauge field as
\begin{equation}
A_t=\frac{\eta\sqrt{3}}{2}\frac{1}{H(r)},
\label{gauge5}
\end{equation}%
where $\eta=+1$ or $\eta=-1$. The Atiyah-Hitchin metric $ds_{AH}^{2}$ is given by the following manifestly 
$SO(3)$ invariant form \cite{GM} 
\begin{equation}
ds_{AH}^{2}=f^{2}(r)dr^{2}+a^{2}(r)\sigma _{1}^{2}+b^{2}(r)\sigma
_{2}^{2}+c^{2}(r)\sigma _{3}^{2},  \label{AHmetric}
\end{equation}%
with 
\begin{eqnarray}
\sigma _{1}&=&-\sin \psi d\theta +\cos \psi \sin \theta d\phi, \label{mcFORMS1}\\ 
\sigma _{2}&=&\cos \psi d\theta +\sin \psi \sin \theta d\phi, \label{mcFORMS2}\\ 
\sigma _{3}&=&d\psi +\cos \theta d\phi, \label{mcFORMS3}
\end{eqnarray}%
where $\sigma _{i\text{ }}$ are Maurer-Cartan one-forms with the property 
\begin{equation}
d\sigma _{i}=\frac{1}{2}\varepsilon _{ijk}\sigma _{j}\wedge \sigma _{k}.
\label{dsigma}
\end{equation}%
We note that the metric on the $\mathbb{R}^{4}$ (with a radial coordinate $R$
and Euler angles ($\theta ,\phi ,\psi $) on an $S^{3}$) could be written in
terms of Maurer-Cartan one-forms by 
\begin{equation}
ds^{2}=dR^{2}+\frac{R^{2}}{4}(\sigma _{1}^{2}+\sigma _{2}^{2}+\sigma
_{3}^{2}).  \label{s3METRIC}
\end{equation}%
We also note that $\sigma _{1}^{2}+\sigma _{2}^{2}$ is the standard metric
of the round unit radius $S^{2}$ and $4(\sigma _{1}^{2}+\sigma
_{2}^{2}+\sigma _{3}^{2})$ gives the same for $S^{3}.$ The metric (\ref%
{AHmetric}) satisfies Einstein's equations provided%
\begin{eqnarray}
a^{\prime }&=&f\frac{(b-c)^{2}-a^{2}}{2bc}, \label{conditions1}\\ 
b^{\prime }&=&f\frac{(c-a)^{2}-b^{2}}{2ca}, \label{conditions2}\\ 
c^{\prime }&=&f\frac{(a-b)^{2}-c^{2}}{2ab}. \label{conditions3}
\end{eqnarray}%
Choosing\textbf{\ }$f(r)=-\frac{b(r)}{r}$\textbf{\ }the explicit expressions
for the metric functions $a,b$ and $c$ are given by%
\begin{eqnarray}
a(r)&=&\sqrt{\frac{r\Upsilon \sin (\gamma )\{\frac{1-\cos (\gamma )}{2}r-\sin
(\gamma )\Upsilon \}}{\Upsilon \sin (\gamma )+r\cos ^{2}(\frac{\gamma }{2})}},
\label{abc1}\\ 
b(r)&=&\sqrt{\frac{\{\Upsilon \sin (\gamma )-\frac{1-\cos \gamma }{2}%
r\}r\{-\Upsilon \sin (\gamma )-\frac{1+\cos \gamma }{2}r\}}{\Upsilon \sin
(\gamma )}}, \label{abc2}\\ 
c(r)&=&-\sqrt{\frac{r\Upsilon \sin (\gamma )\{\frac{1+\cos (\gamma )}{2}r+\sin
(\gamma )\Upsilon \}}{-\Upsilon \sin (\gamma )+\frac{1-\cos \gamma }{2}r}},\label{abc3}
\end{eqnarray}%
where 
\begin{equation}
\Upsilon =\frac{2nE\{\sin (\frac{\gamma }{2})\}}{\sin (\gamma )}-\frac{%
nK\{\sin (\frac{\gamma }{2})\}\cos (\frac{\gamma }{2})}{\sin (\frac{\gamma }{%
2})},  \label{GAMMA}
\end{equation}%
and 
\begin{equation}
K(\sin (\frac{\gamma }{2}))=\frac{r}{2n}.  \label{gama}
\end{equation}%
In the above equations, $K$ and $E$ are the elliptic integrals 
\begin{eqnarray}
K(k) &=&\int_{0}^{1}\frac{dt}{\sqrt{1-t^{2}}\sqrt{1-k^{2}t^{2}}}%
=\int_{0}^{\pi /2}\frac{d\theta }{\sqrt{1-k^{2}\cos ^{2}\theta }},
\label{Ell} \\
E(k) &=&\int_{0}^{1}\frac{\sqrt{1-k^{2}t^{2}}dt}{\sqrt{1-t^{2}}}%
=\int_{0}^{\pi /2}\sqrt{1-k^{2}\cos ^{2}\theta }d\theta,
\end{eqnarray}%
and the coordinate $r$ ranges over the interval $[n\pi ,\infty )$, which
corresponds to $\gamma \in \lbrack 0,\pi ).$ The positive number $n$\ is a
constant number with unit of length that is related to NUT charge of metric
at infinity obtained from Atiyah-Hitchin metric (\ref{AHmetric}).

In fact as $r\rightarrow \infty ,$ the metric (\ref{AHmetric}) reduces to 
\begin{equation}
ds_{AH}^{2}\rightarrow (1-\frac{2n}{r})(dr^{2}+r^{2}d\theta ^{2}+r^{2}\sin
^{2}\theta d\phi ^{2})+4n^{2}(1-\frac{2n}{r})^{-1}(d\psi +\cos \theta d\phi
)^{2},  \label{reducedAH}
\end{equation}%
which is the well known Euclidean Taub-NUT metric with a negative NUT charge 
$N=-n.$ 
The metric (\ref{reducedAH}) is obtained from a consideration of the limiting behaviors
of the functions $a,b$ and $c$ at large monopole separation that are given by%
\begin{eqnarray}
a(r)&=&r(1-\frac{2n}{r})^{1/2}+O(e^{-r/n}), \label{abcatinfinity1}\\ 
b(r)&=&r(1-\frac{2n}{r})^{1/2}+O(e^{-r/n}), \label{abcatinfinity2}\\ 
c(r)&=&-2n(1-\frac{2n}{r})^{-1/2}+O(e^{-r/n}). \label{abcatinfinity3}%
\end{eqnarray}%
In the other extreme limit where $\epsilon=r-n\pi\rightarrow 0$, from equations (\ref{abc1}-\ref{abc3}), we find the following behaviors for the metric functions $a(r), b(r)$ and $c(r)$ 
\begin{eqnarray}
a(r)&=&2\epsilon+O(\epsilon^2),\label{anp}\\
b(r)&=&n\pi+\frac{\epsilon}{2}+O(\epsilon^2),\label{bnp}\\
c(r)&=&-n\pi+\frac{\epsilon}{2}+O(\epsilon^2).\label{cnp}
\end{eqnarray}
Equation (\ref{anp}) shows clearly a bolt singularity as
$\epsilon \rightarrow 0$. Actually, 
by
using the $SO(3)$ invariance of the metric, we can write the metric element (%
\ref{AHmetric}) near the bolt location as%
\begin{equation}
ds^{2}=d\epsilon^{2}+4\epsilon^{2}(d\widetilde{\psi }+\cos \widetilde{\theta }d%
\widetilde{\phi })^{2}+\pi ^{2}n^{2}(d\widetilde{\theta }+\sin ^{2}%
\widetilde{\theta }d\widetilde{\phi }),  \label{nearbolt}
\end{equation}%
where $\widetilde{\psi },\widetilde{\theta }$ and $\widetilde{\phi }$ are
a new set of Euler angles related to $\psi ,\theta ,$\ $\phi $\ by%
\begin{equation}
\mathcal{R}_{1}(\widetilde{\psi })\mathcal{R}_{3}(\widetilde{\theta })%
\mathcal{R}_{1}(\widetilde{\phi })=\mathcal{R}_{3}(\psi )\mathcal{R}%
_{2}(\theta )\mathcal{R}_{3}(\phi ),  \label{Euler}
\end{equation}%
in which $R_{i}(\alpha )$\ represents a rotation by $\alpha $\ about the $i$-
th axis. We note that the last term in (\ref{nearbolt}) is the induced metric
on the two-dimensional bolt.
\section{Einstein-Maxwell Solutions over Atiyah-Hitchin Base Space}
\label{sec:sol}
To find the five-dimensional metric function $H(r)$, we consider equations of motion (\ref{EEQ}-\ref{GEQ}). 
The metric (\ref{ds5}) (along with the gauge field (\ref{gauge5})) is a solution to Einstein-Maxwell equations provided $H\left(r\right)$ is a solution to the
differential equation,
\begin{equation}
ra(r)c(r)\frac{d^2H(r)}{dr^2}+(a(r)b(r)+a(r)c(r)+b(r)c(r)-b(r)^2)\frac{dH(r)}{dr}=0.
\label{eqforH}
\end{equation}
So, we find
\begin{equation}
H(r)=H_0+H_1\int dr e^{\int \frac{b(r)^2-a(r)b(r)-a(r)c(r)-b(r)c(r)}{ra(r)c(r)} dr},
\label{Hsol}
\end{equation}
where $H_0$ and $H_1$ are two constants of integration. Although the $r$-dependences of metric functions $a,b,c$, are given explicitly in 
equations (\ref{abc1}-\ref{abc3}), but it's unlikely to find an analytic expression
for $H(r)$ given by (\ref{Hsol}). 
As $r\rightarrow \infty$, the metric function (\ref{Hsol}) goes to
\begin{equation}
H(r)=H_0-\frac{H_1}{r},
\label{Hatinf}
\end{equation}
On the other hand, near bolt, the metric function $H(r)$ has a logarithmic divergence as 
\begin{equation}
H(r)\simeq \frac{H_1}{4n^2\pi^2}\ln(\epsilon)+H_0 + O(\epsilon),
\label{Honbolt}
\end{equation}
where $\epsilon=r-n\pi$. This type of divergence in the metric function has been observed previously in the metric function of M2-brane in a transverse Atiyah-Hitchin space \cite{GM}.

\begin{figure}[tbp]
\centering           
\begin{minipage}[c]{.3\textwidth}
        \centering
        \includegraphics[width=\textwidth]{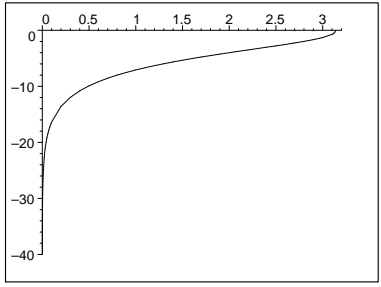}
    \end{minipage}
\caption{
The coordinate $0 \leq \vartheta \leq \pi$ versus coordinate $-\infty < \xi \leq 0$. 
}
\label{xivertheta}
\end{figure}

\begin{figure}[tbp]
\centering           
\begin{minipage}[c]{.3\textwidth}
        \centering
        \includegraphics[width=\textwidth]{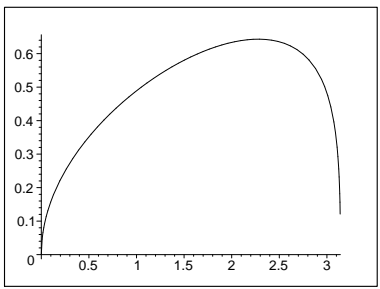}
    \end{minipage}
\caption{
The function $\mu$ versus $\vartheta$. 
}
\label{vartheta}
\end{figure}

\begin{figure}[tbp]
\centering           
\begin{minipage}[c]{.3\textwidth}
        \centering
        \includegraphics[width=\textwidth]{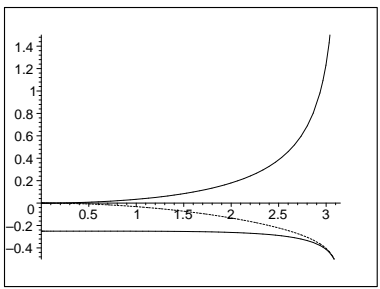}
    \end{minipage}
\caption{
The functions $\psi_1$ (solid bottom), $\psi_2$ (dashed) and $\psi_3$ (solid top) plotted as functions of $\vartheta$. 
}
\label{psis}
\end{figure}

\begin{figure}[tbp]
\centering           
\begin{minipage}[c]{.3\textwidth}
        \centering
        \includegraphics[width=\textwidth]{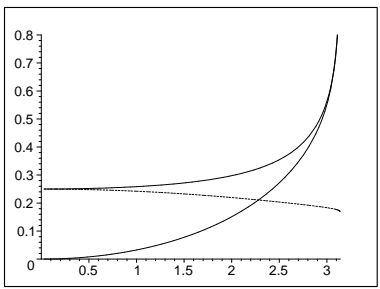}
    \end{minipage}
\caption{
The functions $a$ (solid bottom), $b$ (solid top) and $c$ (dashed) plotted as functions of $\vartheta$. 
}
\label{abcs}
\end{figure}

As we noticed, we could not find a closed analytic expression for the metric function given 
in (\ref{Hsol}). 
To overcome this problem, we choose $f(\xi)$ in (\ref{AHmetric}) to be
$16a(\xi)b(\xi)c(\xi)$ and so the Atiyah-Hitchin metric reads
\begin{equation}
ds_{AH}^{2}=16a^{2}(\xi)b^{2}(\xi)c^{2}(\xi)d\xi^{2}+a^{2}(\xi)\sigma _{1}^{2}+b^{2}(\xi)\sigma
_{2}^{2}+c^{2}(\xi)\sigma _{3}^{2},  \label{AHmetric2}
\end{equation}
where functions $a(\xi),b(\xi)$ and $c(\xi)$ satisfy equations (\ref{conditions1}-\ref{conditions3}) with $f(\xi)=4a(\xi)b(\xi)c(\xi)$
and $'$ means $\frac{d}{d\xi}$. By introducing the new functions $\psi_1(\xi),\psi_2(\xi)$ and $\psi_3(\xi)$ such that 
\begin{eqnarray}
a^2(\xi)&=&\frac{\psi_2\psi_3}{4\psi_1}\label{psis1},\\
b^2(\xi)&=&\frac{\psi_3\psi_1}{4\psi_2}\label{psis2},\\
c^2(\xi)&=&\frac{\psi_1\psi_2}{4\psi_3}\label{psis3},
\end{eqnarray}
the set of equations (\ref{conditions1}-\ref{conditions3}) with $f(\xi)=4a(\xi)b(\xi)c(\xi)$ reduces onto a Darboux-Halpern system
\begin{eqnarray}
\frac{d}{d\xi}(\psi_1+\psi_2)+2\psi_1\psi_2&=&0,\label{H1}\\
\frac{d}{d\xi}(\psi_2+\psi_3)+2\psi_2\psi_3&=&0,\label{H2}\\
\frac{d}{d\xi}(\psi_3+\psi_1)+2\psi_3\psi_1&=&0.\label{H3}
\end{eqnarray} 
We can find the solutions to the above equations as 
\begin{eqnarray}
\psi_1&=&-\frac{1}{2}(\frac{d}{d\vartheta}\mu ^2+\frac{\mu ^2}{\sin\vartheta}),
\label{psi1} \\
\psi_2&=&-\frac{1}{2}(\frac{d}{d\vartheta}\mu ^2-\frac{\mu ^2\cos\vartheta }{\sin\vartheta}),
\label{psi2} \\
\psi_3&=&-\frac{1}{2}(\frac{d}{d\vartheta}\mu ^2-\frac{\mu ^2}{\sin\vartheta}),
\label{psi3} 
\end{eqnarray}
where 
\begin{equation}
\mu (\vartheta)=\frac{1}{\pi}\sqrt{\sin\vartheta}K(\sin\frac{\vartheta}{2}). \label{varw}
\end{equation}
The new coordinate $\vartheta$ is related to the coordinate $\xi$ by 
\begin{equation}
\xi=-\int _\vartheta ^ \pi \frac{d\vartheta}{\mu ^2 (\vartheta)}. 
\label{theta}
\end{equation}
Figure (\ref{xivertheta}) shows result of numerical integration the relation between two coordinates $\xi$ and $\vartheta$. The coordinate $\vartheta$ takes values over $[0,\pi]$ if 
the coordinate $\xi$ is chosen to take values on $(-\infty,0]$. 
In figure (\ref{vartheta}), function $\mu (\vartheta)$ is plotted which shows
an increasing behavior from $\vartheta=0$ to $\vartheta_0=2.281318$. At 
$\vartheta=\vartheta_0$, the
function $\mu$ reaches to maximum value $0.643243$ and decreases then to zero at $\vartheta=\pi$. Hence, in the range of $0 < \vartheta <\pi$, $\mu$ is positive and so
the change of variables, given in (\ref{theta}), is completely well defined.
As one can see from figure (\ref{psis}), functions $\psi_1,\psi_2$ are always negative and $\psi_3$ is always positive. 
Hence, equations (\ref{psis1}, \ref{psis2}, \ref{psis3}) show the metric functions always are positive. In figure (\ref{abcs}), the behaviors of functions $a,b$ and $c$ versus $\vartheta$ are plotted.

The five-dimensional metric and gauge field are given by
\begin{equation}
ds^2=-\frac{dt^2}{(\alpha\xi+\beta)^2}+(\alpha\xi+\beta)\{16a^{2}(\xi)b^{2}(\xi)c^{2}(\xi)d\xi^{2}+a^{2}(\xi)\sigma _{1}^{2}+b^{2}(\xi)\sigma
_{2}^{2}+c^{2}(\xi)\sigma _{3}^{2}\},\label{AH5ex}
\end{equation}
and
\begin{equation}
A_t=\frac{\eta\sqrt{3}}{2(\alpha\xi+\beta)}.\label{gaugefin}
\end{equation}
To avoid any singularity at a finite $\xi$, the sign of $\alpha$ must be opposite to the sign of $\beta$. Moreover, to get a regular positive definite metric, we should choose $\alpha < 0$ and $\beta > 0$. The geometry of solution on a $t=$constant hyper-surface is quite simple. At $\xi=0$ which corresponds to $\vartheta=\pi$, the metric functions given in equations (\ref{psis1}, \ref{psis2}, \ref{psis3}) reduce to
\begin{eqnarray}
a=b&\simeq& \frac{-1}{2\pi}\ln\epsilon+\frac{\ln 8}{2\pi},    \label{aandbatxi0}\\
c&\simeq&\frac{1}{2\pi},\label{catxi0}
\end{eqnarray}
where $\epsilon=\pi-\vartheta$. Hence, we get
\begin{equation}
ds^2\text{\textbar}_{t=cons.} \sim  \frac{-\alpha}{4\zeta}\{d\zeta^2+\zeta^2(d\theta^2+\sin^2\theta d\phi^2)+(d\psi+\cos\theta d\phi)^2 \},
\end{equation}
which is conformaly the Euclidean Taub-NUT metric \cite{GH} or fibration of a unit circle (parametrized with $\psi$) over $R^3$. Here the coordinate $\zeta$ is related to $\epsilon$ by $\zeta=-\ln\epsilon$ and to $\xi$ by $\zeta=-\frac{\pi^2}{\xi}$, respectively. The Ricci scalar of the spacetime (\ref{AH5ex}) approaches $f(\alpha,\beta)\xi ^4(1+O(\xi))$ as $\xi \rightarrow 0$ and the Kretschman invariant approaches $g(\alpha,\beta)\xi ^6(1+O(\xi))$ where $f,g$ are functions of $\alpha$ and $\beta$. On the other extreme level where $\xi \rightarrow -\infty$, which corresponds to $\vartheta \rightarrow 0$, the metric functions behave as
\begin{eqnarray}
a&\simeq&\frac{\vartheta ^2}{768}(24+\vartheta^2+O(\vartheta^4)),   \label{aatxiinf}\\
b&\simeq&\frac{1}{4}(1+\frac{\vartheta ^2}{32}+O(\vartheta^4)),   \label{batxiinf}\\
c&\simeq&\frac{1}{4}(1-\frac{\vartheta ^2}{32}+O(\vartheta^4)).   \label{catxiinf}
\end{eqnarray}
In this limit, the Elliptic integral in equation (\ref{varw}) approaches
\begin{equation}
K(\sin\frac{\vartheta}{2})\simeq \frac{\pi}{2}(1+\frac{1}{16}\vartheta^2+
O(\vartheta^4)),
\end{equation}
hence from equations (\ref{varw}) and (\ref{theta}), we get
\begin{equation}
\xi\simeq 4\ln \vartheta, 
\end{equation}
and we find the metric as
\begin{equation}
ds^2\text{\textbar}_{t=cons.} \sim 4\alpha \ln \vartheta \{
(\frac{\vartheta}{32})^2 d\vartheta^2 + (\frac{\vartheta ^2}{32})^2
\sigma _1^2+\frac{1}{16}(\sigma_2^2+\sigma_3^2)\}.\label{bolt}
\end{equation}
By changing to the coordinate $\varrho=\frac{\vartheta^2}{32}$, the metric (\ref{bolt})  changes to
\begin{equation}
ds^2\text{\textbar}_{t=cons.} \sim 
d\varrho^2 + 4\varrho ^2 \sigma _1^2+\frac{1}{4}(\sigma_2^2+\sigma_3^2),\label{bolt2}
\end{equation}
(up to a conformal factor) which clearly shows a bolt at $\vartheta=0$ of fixed radius $1/2$. The Ricci scalar and Kretschman invariant of the metric (\ref{AH5ex}) near the bolt behave as 
$\frac{(\ln\vartheta)^3}{\vartheta^4}$ and $\frac{1}{(\ln\vartheta)^6\vartheta^8}$ respectively. 
One way to avoid bolt region is to consider positive values for both $\alpha$ and $\beta$.  In this case, the range of $\xi$ is limited to $\xi _0 \leq \xi \leq 0$ where $\xi _0=-\frac{\beta}{\alpha}$. Although there is still a curvature singularity at $\xi=\xi _0$ of the order of $\frac{1}{\varepsilon ^3}$ where $\varepsilon=\xi-\xi_0$, but it is quite less divergent than singularity on the bolt. This latter singularity is a simple result of our symmetric metric function $H(r)$ in the ansatz (\ref{ds5}) that could be removed by considering some non-symmetric metric functions. we leave this case along with some other open issues for a future article.

\section{Concluding Remarks}

\label{sec:con}

The main result of this article is the metric (\ref{AH5ex})
along with the gauge field (\ref{gaugefin}) that are exact solutions to the five-dimensional Einstein-Maxwell equations. To our knowledge, these solutions are the first known solutions to five-dimensional Einstein-Maxwell theory based on non-triholomorphic base space, hence they could be used to study the physical processes that do not have any triholomorphic symmetry. Simplicity of these solutions (simple analytic metric functions) is a result of taking the base Atiyah-Hitchin metric in the form of (\ref{AHmetric2}); otherwise the metric function (\ref{Hsol}) cannot be obtained in a simple analytic form.
The metric function and the gauge field are regular everywhere in spacetime. The metric is regular everywhere except on the location of original bolt in four-dimensional Atiyah-Hitchin space. The similar results have been observed previously in higher-dimensional (super)gravity solutions based on transverse self-dual hyper-K\"{a}hler manifolds \cite{GH, H, GH2}.

We conclude with a few comments about possible directions for future work. 
In our solutions, we have considered the simplest dependence of the five-dimensional metric function and gauge field on the coordinates (i.e. dependence only to the radial coordinate). We can seek for other solutions for which the functions appearing in the metric, depend on more coordinates. 
It's quite possible that in these solutions, the singularity in the location of bolt can be converted to a regular hypersurface(s) in five-dimensional space-time and we obtain Atiyah-Hitchin black hole solutions.
The other possibility is to include the cosmological constant into the theory that may lift the singularity behind some regular hypersurface(s). Moreover the solutions could be used to study (A)dS/CFT correspondence where Atiyah-Hitchin space is a part of bulk spacetime.
The other open issue is study of the thermodynamics of solutions constructed in
this paper.

\bigskip
{\Large Acknowledgments}

This work was supported by the Natural Sciences and Engineering Research
Council of Canada.

\end{document}